%% file: main.tex
\RequirePackage{amsmath}
\documentclass[svgnames,a4paper]{llncs}
\usepackage{subcaption}
\usepackage{graphicx}
\usepackage[T1]{fontenc}
\usepackage{hyperref}
\usepackage{xspace}
\usepackage{listingsVDM}
\usepackage{times}
\usepackage{subfiles}
\usepackage{float}
\usepackage{xspace}
\usepackage{pdflscape}
\newcommand{\W}{\emph{\textsf{W}}\xspace}
\begin{document}
    \title{Modelling Maritime SAR Effective Sweep Widths for Helicopters in VDM}
    \author{Alexander Sulaiman\inst{1} \and Ken Pierce\inst{1}}
    \institute{School of Computing, Newcastle University, United Kingdom \\\email{A.Sulaiman2@ncl.ac.uk,kenneth.pierce@ncl.ac.uk}}
    \maketitle
    \begin{abstract}
        Search and Rescue (SAR) is searching for and providing help to people in danger. In the UK, SAR teams are typically charities with limited resources, and SAR missions are time critical. Search managers need to objectively decide which search assets (e.g. helicopter vs drone) would be better. A key metric in the SAR community is \emph{effective sweep width} (\W), which provides a single measure for a search asset’s ability to detect a specific object in specific environmental conditions. Tables of \W for different search assets are provided in various manuals, such as the International Aeronautical and Maritime SAR (IAMSAR) Manual. However, these tables take years of expensive testing and experience to produce, and no such tables exist for drones. This paper uses the Vienna Development Method (VDM) to build an initial model of \W for a known case (helicopters at sea) with a view to predicting \W tables for drones. The model computes \W for various search object sizes, helicopter altitude and visibility. The results for the model are quite different from the published tables, which shows that the abstraction level is not yet correct, however it produced useful insights and directions for the next steps.
    \end{abstract}
    \section{Introduction}
    \subfile{sections/introduction}
    \section{Background} 
    \subfile{sections/background}
    \section{The VDM Model}
    \subfile{sections/model}

    \section{Results}
    \subfile{sections/results}
    \section{Conclusions and Future Work}
    \subfile{sections/conclusions_and_future_work}
    \section*{Acknowledgements}
    \subfile{sections/acknowledgements}
    \bibliographystyle{splncs03}
    \bibliography{bibs/dan,bibs/effective_sweep_width,bibs/iamsar_manual}
\end{document}

%% file: sections/introduction.tex
\label{sec:intro}

Search and Rescue (SAR) covers the search for persons in distress or danger, and the provision of aid to them. While there are several specialised fields, primarily based on the terrain in which the search is conducted, the general problem of search is similar across these. In essence, a \emph{search manager} is responsible for a search has a number of \emph{search assets}; the search manager must select how best to use these assets to find the missing persons (\emph{mispers}) or objects, based on last known location, search area, local knowledge etc. There are a range of search assets that can be used, e.g. humans, dogs, drones, each with some form of \emph{sensor}, e.g. eyes, noses, cameras. Each of these assets has different characteristics in terms of their ability to search an area within a given period and a given level of success. 

Search managers must be able to make quick decisions on how to deploy their available assets during a search; this depends on being able to quickly quantify and compare available assets. Effective sweep width (\W) is a concept that helps in these decisions by providing a single metric for each assets ability to search in a given set of conditions. \W is a key aspect of \emph{search theory}: it allows diverse search assets to be compared easily in order to support fast and high-quality decisions at critical times. Search manuals, such as the IAMSAR Manual (see Section~\ref{sec:background}) provide tables of \W for different types of assets and conditions. The IAMSAR Manual, for example, provides tables for \W for helicopters at a given height and visibility, with modifiers for known information, such as whether the misper is wearing a high-visibility life jacket. 

Accurate \W tables are vitally important to effective SAR, but since they are produced primarily from empirical studies in the field, they can be extremely expensive to run. SAR teams in the UK are primarily operated by small charities (annual income of less than £1m) and staffed by volunteers, and cannot regularly run field trials to generate new information. The increasing availability of low-cost drones ---typically off-the-shelf quadcopters--- with high-quality cameras has led to interest in their use in civilian SAR. Unfortunately, \W is not well understood for drones, and tables for \W or guidelines for their use. \W tables for helicopters and other search assets required ``many years of experience and testing''~\cite[p.107]{imo2022iamsar} to develop, which is out of reach of civilian SAR teams. 

This leads to the potential of using modelling and simulation to run \emph{virtual field trials} to help develop \W tables for drones. This paper presents some early results in work to explore this possibility. Given limited understanding of search theory, this effort is split into various steps:

\begin{enumerate}
    \item Develop a simple, discrete-event model of a field trial for a known \W table (i.e. maritime helicopter search) to aid understanding of \W and its calculation;  
    \item Refine the model to find the key factors affecting \W; 
    \item Propose a model for generating \W for drones and predict \W; and
    \item Run real field trials and evaluate predicted \W against real \W for drones. 
\end{enumerate}

This paper reports on the first step, and provides suggestions for the second step, and future directions for the third. The results show that the model does indeed need to be refined to better reflect the reality of the search. We select the Vienna Development Method (VDM) for its ease of use and tool support for simulation, include combinatorial testing. Also, given that the \emph{environmental conditions} is a key factor, our expectation is that the abstract, two-dimensional `ocean' will need to be replaced with a high-fidelity environment model, and VDM supports this seamlessly through the Functional Mock-up Interface (FMI) and INTO-CPS tool chain. 

In the remainder of this paper, Section~\ref{sec:background} provides background information on SAR concepts. Section~\ref{sec:model} describes the modelling a lateral range experiment. Section~\ref{sec:results} covers the results and evaluation. Section~\ref{sec:conc} provides some closing remarks.

%% file: sections/background.tex
\label{sec:background}

This section introduces the key concept in SAR called Effective Sweep Width (\W), and the aircraft and maritime SAR manual that provides data against which the results of this paper are compared.  

\subsection{Effective Sweep Width}

Effective search (or sweep) width and sweep width are used synonymously in the literature ~\cite{koopman1946search,koopman1980search,imo2022iamsar}. \W is about quantifying how effectively a specific sensor detects a specific object in specific environmental conditions~\cite{imo2022iamsar} by specifying a single measurement for each sensor by which sensors can be compared. \W is a key concept in ``search theory'', which was developed in World War II for naval warfare by Koopman~\cite{koopman1946search,koopman1980search}.
W is defined as the area under the Lateral Range Curve (LRC)~\cite{koopman1946search,koopman1980search,guard1996theory,cooper2003compatibility,washburn2014search,stone2016optimal} as shown in Equation~\ref{eq:w}. An LRC will be explained using a lateral range experiment~\cite{washburn2014search}.

\begin{equation}
W=\int_{-\infty}^{+\infty}p(x)dx
\label{eq:w}
\end{equation}

\begin{figure}
    \centering
    \includegraphics[width=0.8\textwidth]{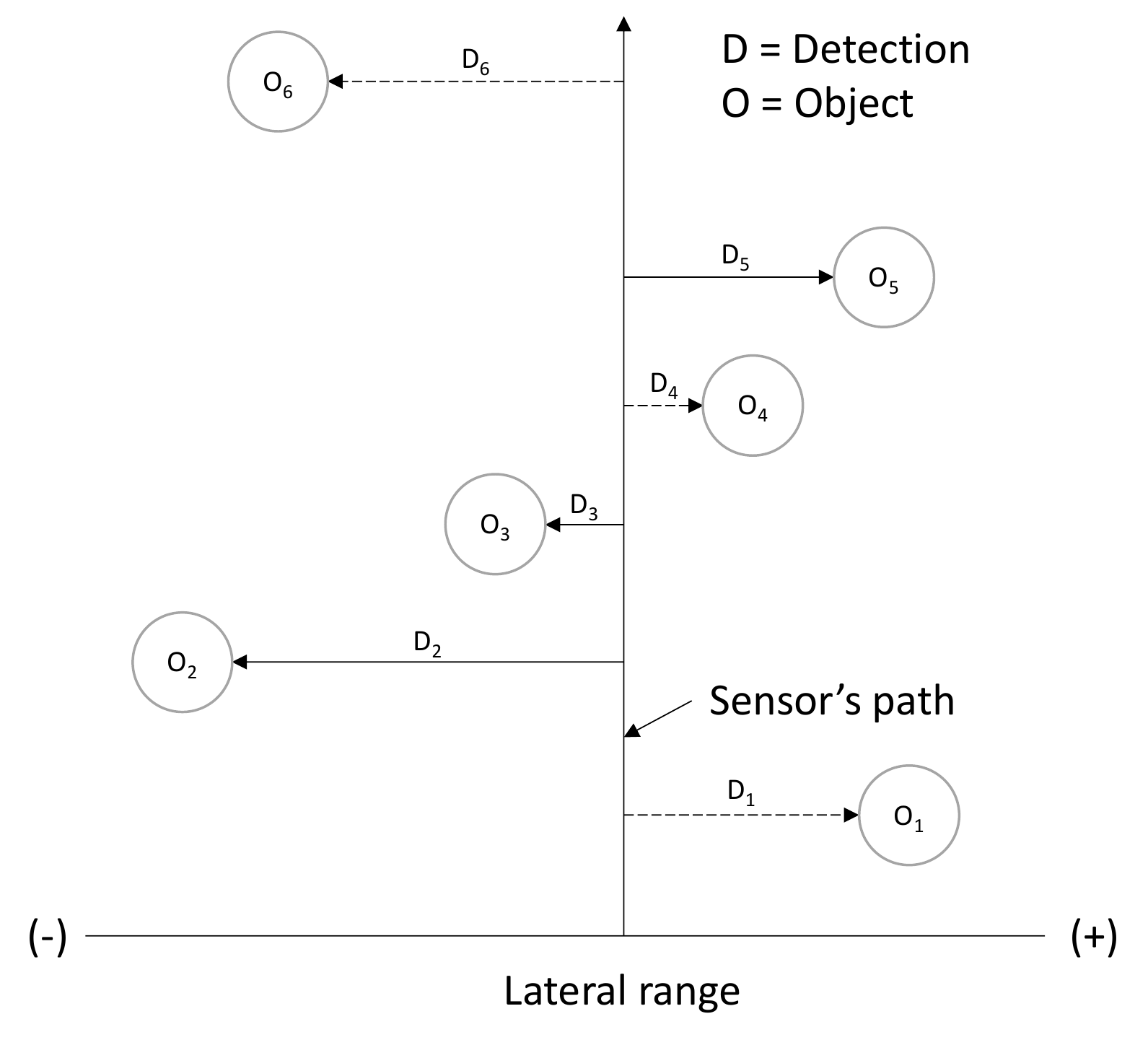}
    \caption{Lateral range experiment}
    \label{fig:lateral_range_experiment}
\end{figure}

The lateral range experiment is represented in Figure~\ref{fig:lateral_range_experiment}. The idea of the lateral range experiment is that a sensor follows a straight path. Along the sensor's path, there are detection opportunities represented as \begin{math}D_1\end{math} to \begin{math}D_6\end{math} for detecting objects \begin{math}O_1\end{math} to \begin{math}O_6\end{math}. The solid arrow represents a detection, and the dashed arrow represents a missed detection. When an object is detected, it is detected at a given lateral range distance. At each lateral range distance, there is detection data for how many objects were detected compared to how many were there, and this is how the LRC is derived. The reason for some objects being detected and some that are not is due to many factors. In the real world, if you think purely from the sensor’s point of view, the sensor may not be perfect and may miss a detection, e.g. if the human was a sensor. The environment could have obstacles that can interfere with the detection. An object's physical properties can cause a missed detection, e.g. if it is too small. Getting a meaningful LRC depends on the detection opportunities: the more, the better, which is hard for a field experiment. The need for many detection opportunities highlight the need for a simulation model. 

\begin{figure}
    \centering
    \includegraphics[width=\textwidth]{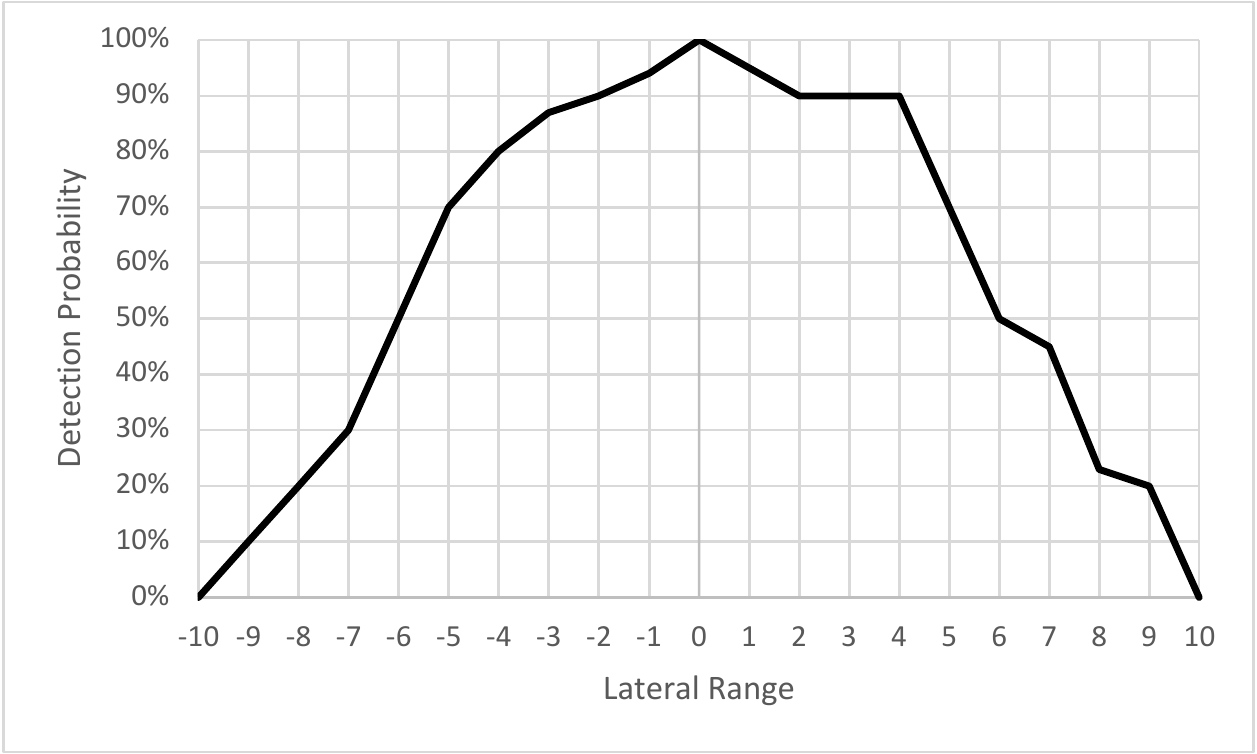}
    \caption{Nonlinear Lateral Range Curve}
    \label{fig:nonlinear_lrc}
\end{figure}

Figure \ref{fig:nonlinear_lrc} represents an example of a sensor's performance profile in reference to its detection capability after a lateral range experiment like Figure~\ref{fig:lateral_range_experiment}. This shows that, at lateral range \begin{math}0\end{math}, \begin{math}100\%\end{math} of objects were detected. At lateral range \begin{math}-9\end{math}, \begin{math}10\%\end{math} of objects were detected. At lateral range \begin{math}6\end{math}, \begin{math}50\%\end{math} of objects were detected, and so on.

\subsection{Related Work}

Values of \W are typically derived from field experiments, e.g. experiments have been conducted for air-scent dog teams~\cite{chiacchia2015deriving} and human visual range detection~\cite{chiacchia2023deriving} for land SAR. Chiacchia \& Houlahan also show that \W is a good predictor of Probability of Detection (PoD)~\cite{chiacchia2023deriving}. The main computer modelling simulation work for \W is by Perkins~\cite{perkins2021a}, who found that drone \W can be modelled by simulating a camera drone missing person search and that a field method can approximate \W. The method they used to get a drone \W is by modelling a camera drone searching a grid by travelling up column zero producing an LRC. \W is then calculated from the area under this curve. The grid consists of targets and obstacles. One target is placed on each row in a random location. Obstacles are placed on each row randomly, which can obstruct the target being detected. The camera drone travels up column zero and detects laterally in each row if the camera drone is capable of detecting and there is not an obstacle in the way. Different variables are modelled, i.e., obstacle density, height, drone height, and camera lens angular field of view. The limitations are that of obstacle clumping and the need for a deeper understanding of obstacle heights to be assessed in how this affects \W. Perkins~\cite{perkins2011some} found that a computer simulation model for a land SAR searcher can be used to quantify detection ranges for targets in different environments. They came to this conclusion by producing an LRC, where the area under the curve can be calculated to produce \W. The limitations are that the searcher never has a decrease in performance, which is not like the real world. Obstacle density could also mean how opaque obstacles are, which is not taken into consideration, and this affects detection for targets.

\subsection{The IAMSAR Manual}

The International Aeronautical and Maritime SAR (IAMSAR) Manual produces guidelines for aircraft and maritime SAR activities. The IAMSAR manual is split into three volumes. Volume I is about the overall SAR system. Volume II is for SAR managers. Volume III is for when on a SAR mission.

The IAMSAR Manual Volume II~\cite{imo2022iamsar} provides a way of measuring the effectiveness of a sensor detecting a search object in given environmental conditions. E.g., sensor (helicopter), search object (ship) and environmental conditions (perfect). The measurement is in the form of \W tables, which are empirically derived.

%% file: sections/model.tex
\label{sec:model}

\begin{figure}
    \centering
    \includegraphics[width=\textwidth]{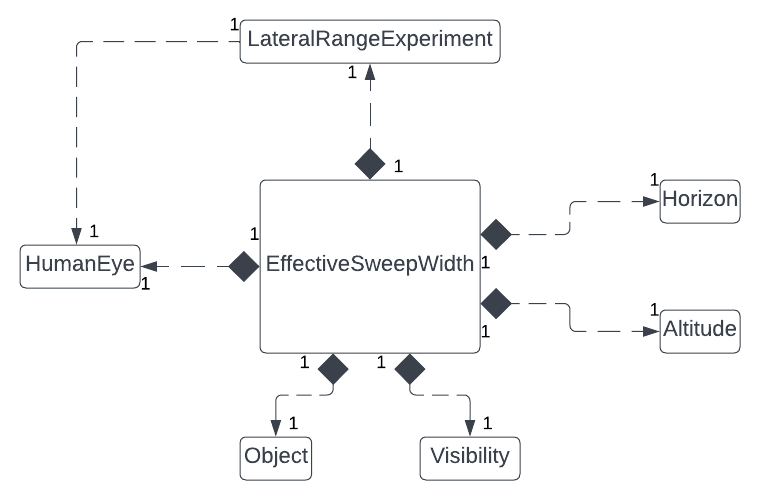}
    \caption{VDM model structure}
    \label{fig:uml_class_diagram}
\end{figure}

This section has been modelled from the IAMSAR Manual Volume II \cite{imo2022iamsar} \W for helicopters N-5 table. Figure \ref{fig:uml_class_diagram} shows the high-level structure of this model as a UML class diagram. The remainder of this section describes the modelling of the sensor, environmental conditions, object, lateral range experiment, and \W. 


\subsection{Sensor}

Listing \ref{lst:altitudes_constant} shows the altitudes constant for the sensor (human eye). The human eye can look from three altitudes (from the helicopter), \begin{math}150\end{math} metres, \begin{math}300\end{math} metres and \begin{math}600\end{math} metres.

\begin{lstlisting}[language=VDM, captionpos=b, caption={Altitudes constant}, label={lst:altitudes_constant}, style=styleVDM]
ALTS_M : seq of nat1 = [150,300,600]
\end{lstlisting}

Listing \ref{lst:human_eye_detection_operation} human eye detection operation is based on the Rayleigh criterion in Equation~\ref{eq:rayleigh_criterion}. In Equation~\ref{eq:rayleigh_criterion}, \begin{math}\lambda\end{math} is the wavelength for visible light, and \begin{math}D\end{math} is the aperture, which is the human eye's pupil diameter. The operation takes in the input altitude (sensor altitude minus object height), object width, and object position. Sets the minimum object size that can be resolved given the inputs and returns whether the object can be detected.

\begin{equation}
\theta\approx1.22\frac{\lambda}{D}
\label{eq:rayleigh_criterion}
\end{equation}

\begin{lstlisting}[language=VDM, frame=single, breaklines, captionpos=b, caption={Human eye detection operation}, label={lst:human_eye_detection_operation}, style=styleVDM]
detect : nat1 * nat1 * nat1 ==> bool
detect (alt, objSize, objColInd) == (
    setMinimumObjectSizeResolution(alt, objSize, objColInd);
    if objSize >= minObjSRes then (
        return true
    );
    return false
)
\end{lstlisting}

\subsection{Environmental Conditions}

Listing \ref{lst:distance_to_horizon_function} shows the distance to horizon function, taken from the IAMSAR Manual Volume II. The distance to horizon function takes altitude in metres as an input, like from listing \ref{lst:altitudes_constant}, and outputs a distance to the horizon in kilometres. This model assumes that objects can not be seen past the horizon.

\begin{lstlisting}[language=VDM, captionpos=b, caption={Distance to horizon function}, label={lst:distance_to_horizon_function}, style=styleVDM]
distanceToHorizonKilometres :  real -> real
distanceToHorizonKilometres(altMetres) == 
  (3.83 * MATH`sqrt(altMetres))
\end{lstlisting}

Listing \ref{lst:visibilities_constant} shows the sensor's visibility constant that the sensor has to operate in when detecting. The last visibility, \begin{math}37\end{math} kilometres, stands for \begin{math}37\end{math} kilometres and greater.

\begin{lstlisting}[language=VDM, captionpos=b, caption={Visibilities constant}, label={lst:visibilities_constant}, style=styleVDM]
VIS_KM : seq of real = [1.9,5.6,9.3,18.5,27.8,37]
\end{lstlisting}

\subsection{Objects}

Listing~\ref{lst:objects_constant} shows the objects represented as a constant. The objects constant structure maps the object's name to its height and width in metres. For example, a ``raft 1-person'' has a  height and width of 1 metre. ``Person 1'' has been commented out as in this model it is treated the same as ``raft 1-person'', and so on with the other search objects.

\begin{lstlisting}[language=VDM, captionpos=b, caption={Objects constant}, label={lst:objects_constant}, style=styleVDM]
OBJS : map seq of char to nat1 = {
    -- Same as raft 1-person
    -- "Person 1" |-> 1,
    "Raft 1-person" |-> 1,
    "Raft 4-person" |-> 4,
    "Raft 6-person" |-> 6,
    "Raft 8-person" |-> 8,
    "Raft 10-person" |-> 10,
    "Raft 15-person" |-> 15,
    "Raft 20-person" |-> 20,
    "Raft 25-person" |-> 25,
    "Power boat 2" |-> 2,
    -- Same as raft 6-person
    -- "Power boat 6" |-> 6,
    -- Same as raft 10-person
    -- "Power boat 10" |-> 10,
    "Power boat 16" |-> 16,
    "Power boat 24" |-> 24,
    "Sail boat 5" |-> 5,
    -- Same as raft 8-person
    -- "Sail boat 8" |-> 8,
    "Sail boat 12" |-> 12,
    -- Same as raft 15-person
    -- "Sail boat 15" |-> 15,
    "Sail boat 21" |-> 21,
    -- Same as raft 25-person
    -- "Sail boat 25" |-> 25,
    "Ship 37" |-> 37,
    "Ship 69" |-> 69,
    "Ship 92" |-> 92
}
\end{lstlisting}

\subsection{Lateral Range Experiment}

\begin{figure}[tb]
    \centering
    \includegraphics[width=\textwidth]{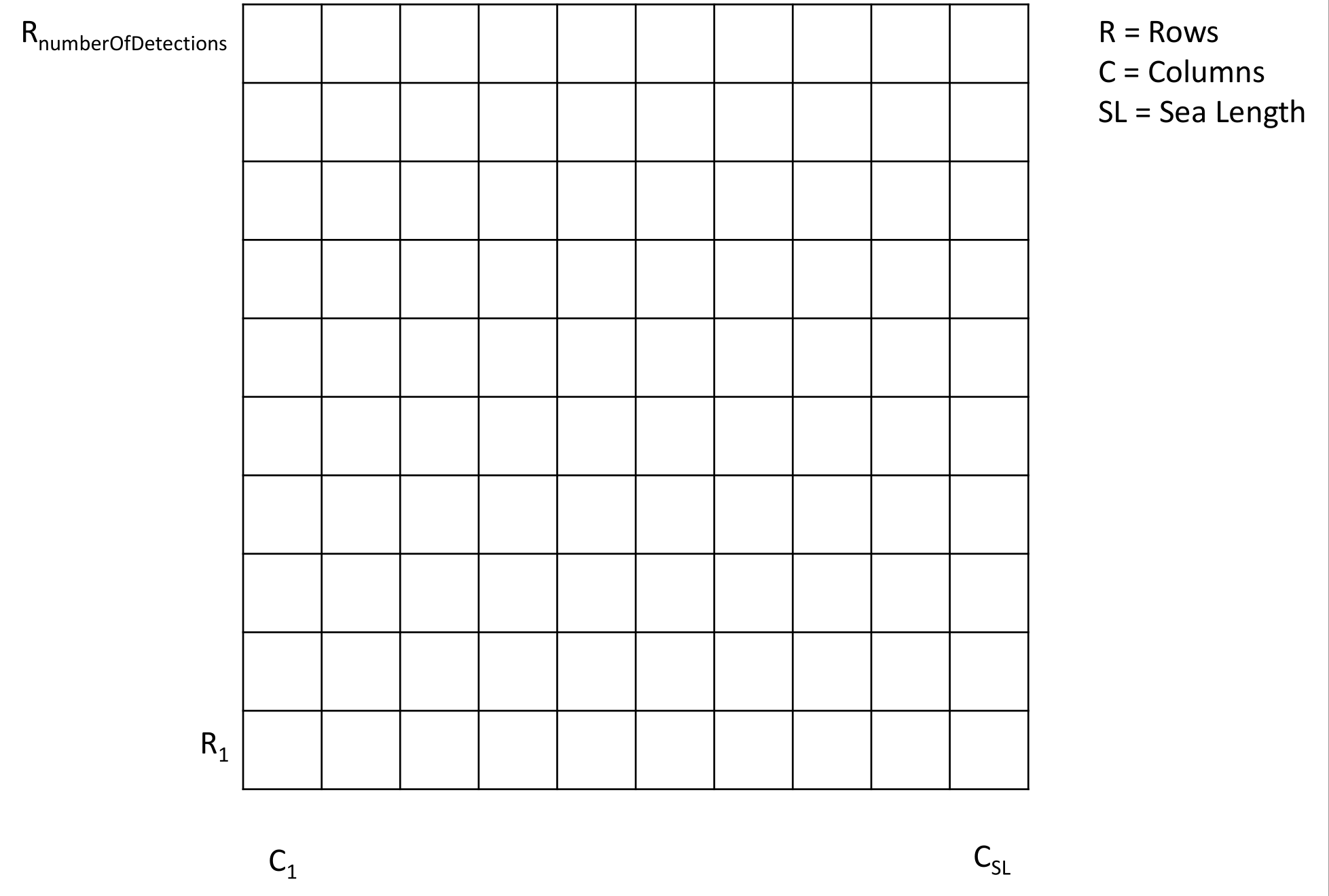}
    \caption{Lateral range experiment grid setup}
    \label{fig:lateral_range_experiment_grid_setup}
\end{figure}

\begin{figure}
    \centering
    \includegraphics[width=\textwidth]{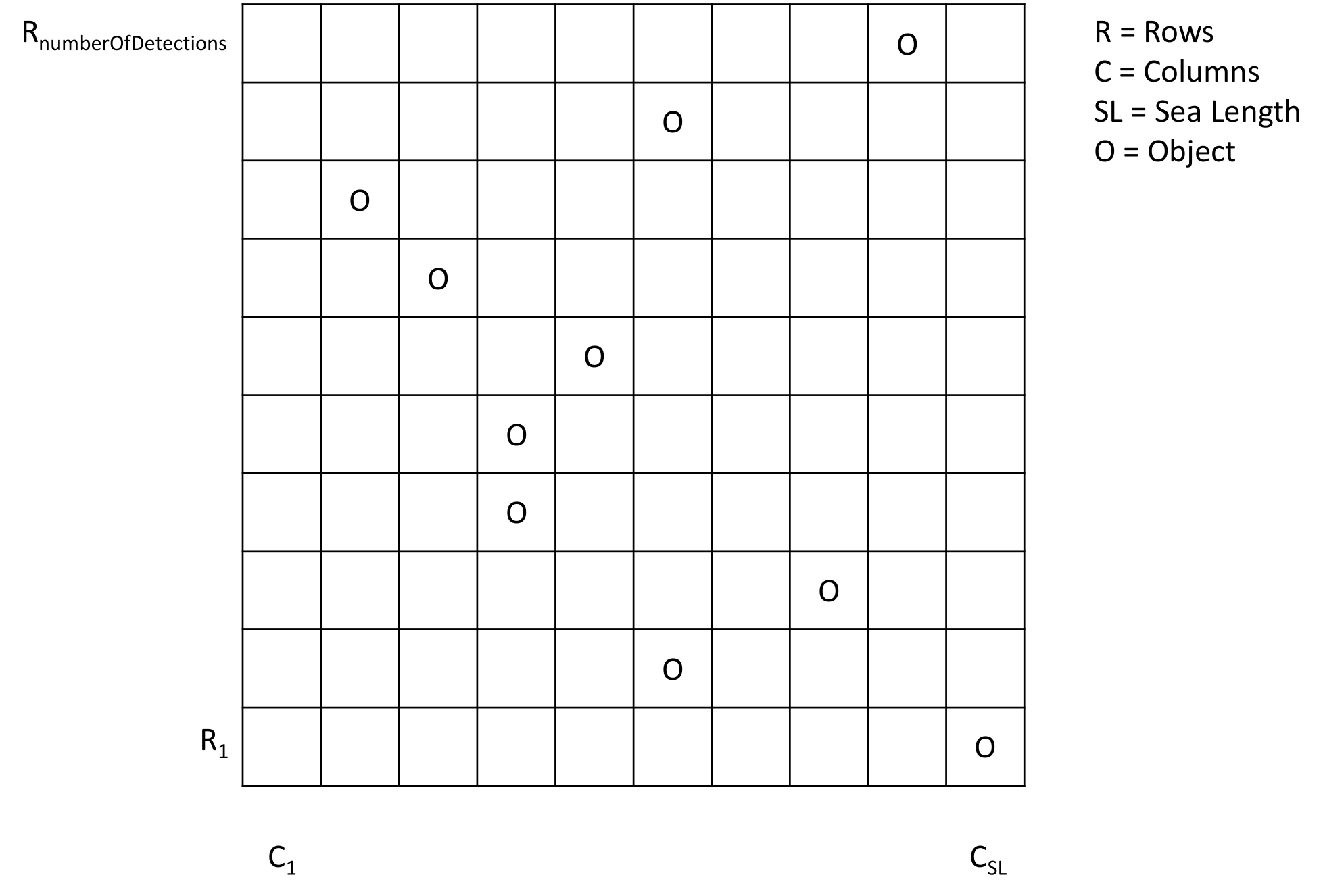}
    \caption{Lateral range experiment randomly place objects}
    \label{fig:lateral_range_experiment_place_search_objects}
\end{figure}

Figure \ref{fig:lateral_range_experiment_grid_setup} shows the lateral range experiment grid setup. The setup consists of a grid of rows and columns. The number of rows determines the maximum number of possible detections. The length of the column length is the sea length defined as \begin{math}54200\end{math} metres. Figure \ref{fig:lateral_range_experiment_place_search_objects} represents the next step in the lateral range experiment. The next step is to place one object in a random location in each row.

\begin{figure}
    \centering
    \includegraphics[width=\textwidth]{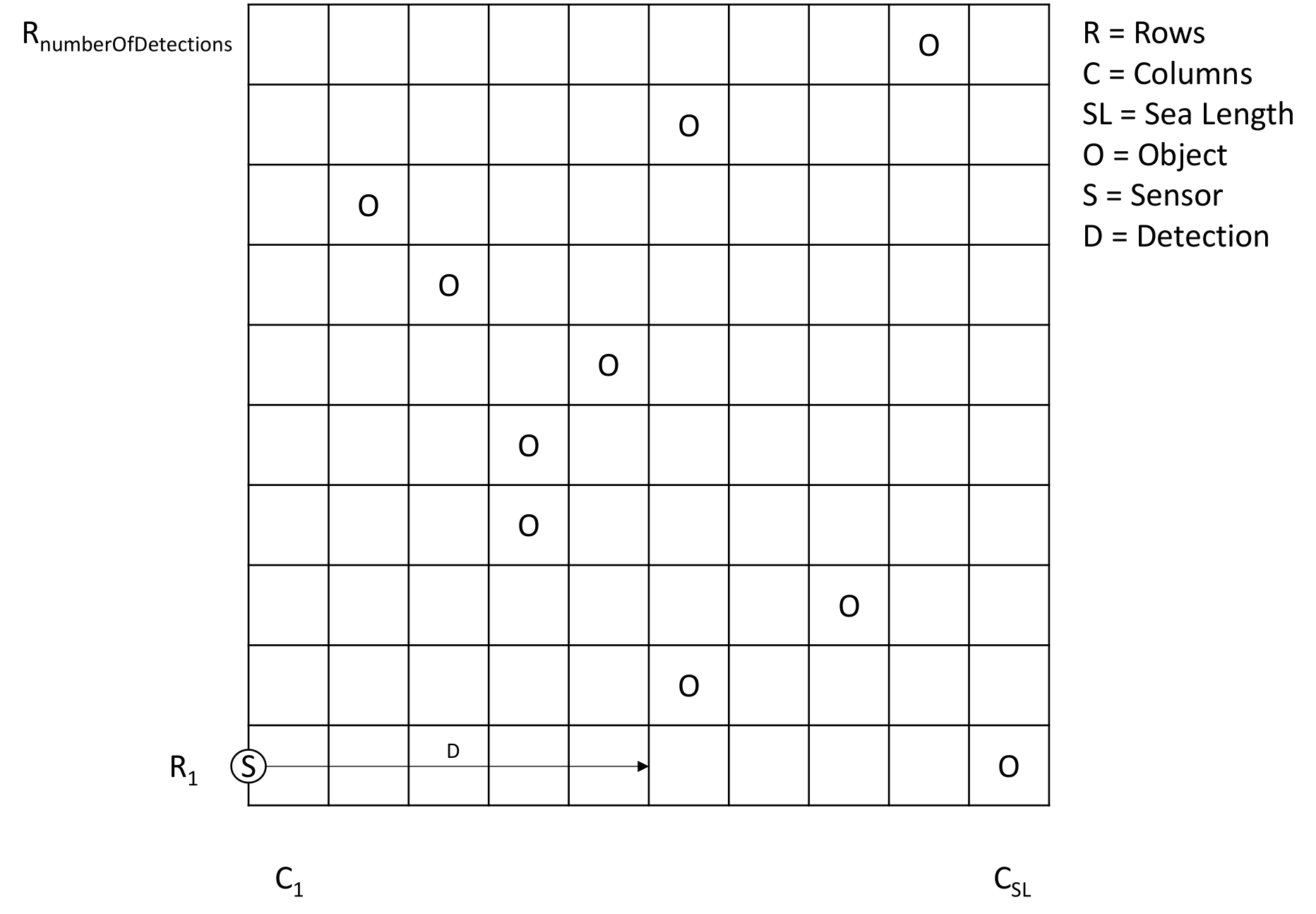}
    \caption{Lateral range experiment detection}
    \label{fig:lateral_range_experiment_detection}
\end{figure}

Figure \ref{fig:lateral_range_experiment_detection} represents the last step of the lateral range experiment. The sensor (human eye in a helicopter) goes along each row and detects laterally depending on Listings~\ref{lst:human_eye_detection_operation}, \ref{lst:distance_to_horizon_function}, and \ref{lst:visibilities_constant}.

The lateral range experiment detection operation is shown in Listing~\ref{lst:lateral_range_detection_experiment_operation}. 
The lateral range experiment detection operation is set up by first converting the distance to the horizon from kilometres from Listing~\ref{lst:distance_to_horizon_function} to metres, then converting visibility from kilometres from Listing~\ref{lst:visibilities_constant} to metres. 

If the object's position is less than or equal to the distance to the horizon, it proceeds, or else returns false for detection, as the assumption is that objects cannot be seen past the horizon. If the object's position is less than or equal to the distance to the horizon, it checks whether visibility is greater than \begin{math}37\end{math} kilometres. If so, the sensor's detect operation from Figure~\ref{lst:human_eye_detection_operation} returns true (no visibility limitations). Otherwise, if visibility is not greater than \begin{math}37\end{math} kilometres, it makes sure the object position is within the visibility, and if so, the sensor's detect operation returns true. 

\newpage
\begin{lstlisting}[language=VDM, captionpos=b, caption={Lateral range experiment detection operation}, label={lst:lateral_range_detection_experiment_operation}, style=styleVDM]
detect: HumanEye * nat1 * nat1 * nat1 * real * real ==> bool
detect(sen, alt, objSize, objColInd, dstToHorizon, vis) == (
    dcl dstToHorizM : real := dstToHorizon * 1000;
    dcl visM : real := vis * 1000;
    dcl thirtySevenKmGreaterCase : nat1 := 37000;
    if objColInd <= dstToHorizM then (
        if visM = thirtySevenKmGreaterCase then (
            return sen.detect(alt, objSize, objColInd)
        ) else if objColInd <= vis then (
            return sen.detect(alt, objSize, objColInd)
        )
    );
    return false
)
\end{lstlisting}

Listing \ref{lst:object_detection_data_structure} represents that each object has detection data based on its column position mapped to a tuple consisting of how many times the object has been detected, how many detection opportunities there are, and the percentage detected.

\begin{lstlisting}[language=VDM, captionpos=b, caption={Object detection data structure}, label={lst:object_detection_data_structure}, style=styleVDM]
objDetnD : map nat1 to (nat * nat * real)
\end{lstlisting}

\subsection{Effective Sweep Width}

Listing \ref{lst:main_effective_sweep_width_operation} represents the main sweep width operation containing three for-loops. The outer loop is the objects in Listing~\ref{lst:objects_constant}, the middle loop is the sensor (human eye in a helicopter) altitudes in Listing~\ref{lst:altitudes_constant}, and the inner loop is visibilities in Listing~\ref{lst:visibilities_constant}. The distance to the horizon in Listing~\ref{lst:distance_to_horizon_function} is calculated using altitude in the middle loop. In the inner loop, the lateral range experiment is executed, which produces object detection data based on object size, altitude, visibility and distance to the horizon given by the sensor. Lastly, the object detection data is used to calculate the \W.

\begin{lstlisting}[language=VDM, captionpos=b, caption={Main effective sweep width operation}, label={lst:main_effective_sweep_width_operation}, style=styleVDM]
main () ==
    for all objName in set dom obj.getObjects() do (
        for alt in alt.getAltitudes() do (
            horiz.setDistance(altitude);
            for v in vis.getVisibilities() do (
                dcl objS : nat1 := obj.getObjects()(objName);
                lre.run(sen, alt, objS, horiz.getDistance(), v);
                calculateW()
            )
        )
    )
\end{lstlisting}

Listing~\ref{lst:calculate_effective_sweep_width_operation} represents how \W is calculated. \W is calculated by getting the object detection data for all the column index positions the object was placed in from the lateral range experiment shown in Figure~\ref{fig:lateral_range_experiment_place_search_objects}. Then total up the percentage detected for each object column index position shown in Listing~\ref{lst:object_detection_data_structure}. The total is then converted from metres to kilometres, and multiplied by two as currently, the lateral range experiment calculated the right side of the LRC for the sensor detecting the object. The assumption for this model is that the LRC is symmetrical, not nonlinear, like in Figure~\ref{fig:nonlinear_lrc}.

\begin{lstlisting}[language=VDM, caption={Calculate effective sweep width operation}, label={lst:calculate_effective_sweep_width_operation}, style=styleVDM]
calculateW: () ==> ()
calculateW() == (
    dcl wM : real := 0;
    dcl objDetnD : map nat to (nat * nat * real) :=
        lre.getObjectDetectionData();
    dcl kmToM : nat1 := 1000;
    dcl symmetricalLRC := 2;
    for all objDetnDPos in set dom objDetnD do wM := 
        wM + objDetnD(objDetnDPos).#3;
    w := wM / kmToM;
    w := w * symmetricalLRC;
)
\end{lstlisting}

%% file: sections/results.tex
\label{sec:results}

This section discusses the results produced from the simulation model; this is a table of \W for human visual detection in a helicopter at sea based on different object sizes, altitudes, and visibility. It then discusses the absolute difference between the \W table from this simulation model results and the expected results from the IAMSAR Manual Volume II \W for helicopters. Lastly, it focuses on the 600 metres altitude case for the simulation model and the IAMSAR manual \W table described using a three-dimensional scatterplot to evaluate the results more closely.

\begin{figure}
    \centering
    \includegraphics[angle=90, height=\textheight]{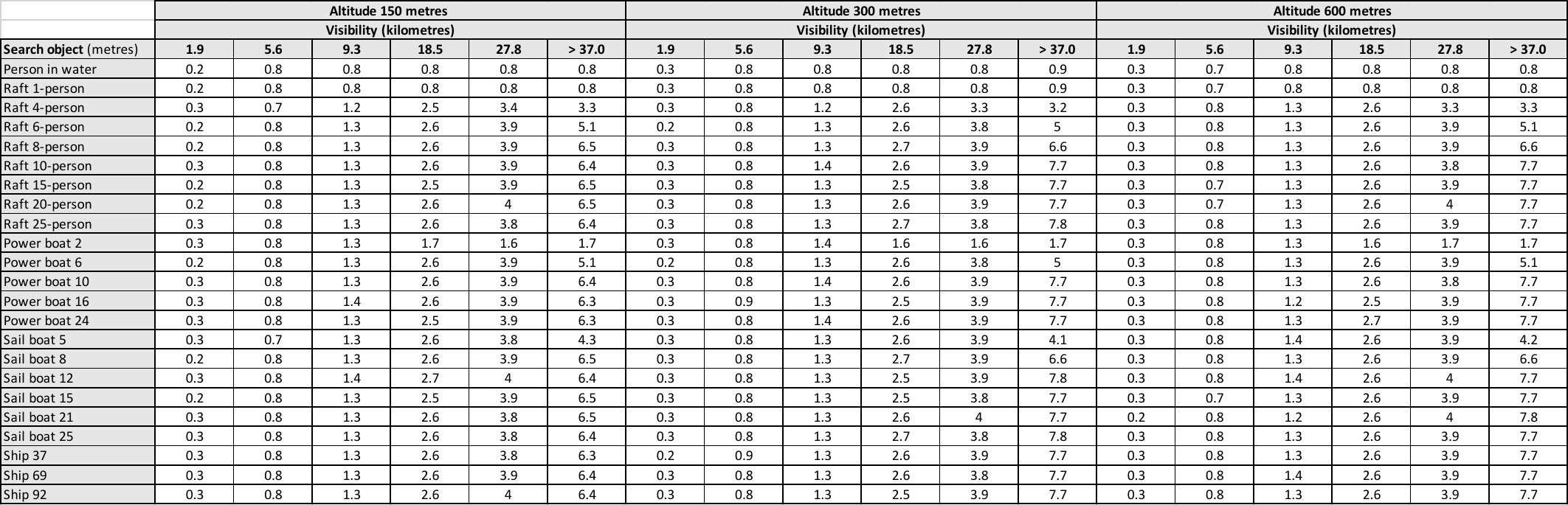}
    \caption{Simulation model effective sweep widths table 4000 human visual detections for each object in a helicopter at sea result}
    \label{fig:w_actual_table}
\end{figure}

\begin{figure}
    \centering
    \includegraphics[angle=90, height=\textheight]{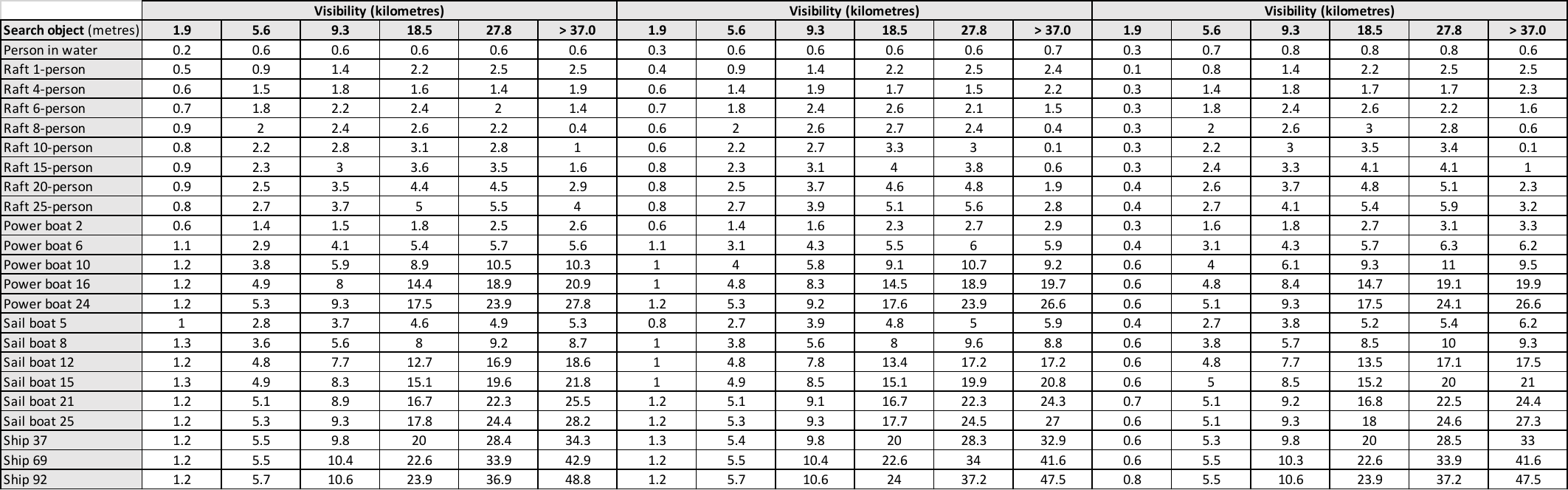}
    \caption{Effective sweep widths table absolute difference between the simulation model 4000 human visual detections for each object in a helicopter at sea and the IAMSAR manual sweep widths for helicopters results}
    \label{fig:w_table_absolute_difference_actual_vs_expected}
\end{figure}

Figure \ref{fig:w_actual_table} shows the simulation model results from section \ref{sec:model} for \W calculation for \begin{math}4000\end{math} human visual detections for each object in a helicopter at sea at different altitudes and visibilities. E.g., the \W for detecting a ``Raft 8-person'' search object at \begin{math}300\end{math} metres altitude with \begin{math}9.3\end{math} kilometres visibility is \begin{math}1.3\end{math} kilometres. Figure \ref{fig:w_actual_table} shows the same \W for objects ``Person in water'' and ``Raft 1-person'', ``Power boat 6'' and ``Raft 6-person'', ``Power boat 10'' and ``Raft 10-person'', ``Sail boat 8'' and ``Raft 8-person'', ``Sail boat 15'' and ``Raft 15-person'', ``Sail boat 25'' and ``Raft 25-person'' as this is how the objects have been modelled. Figure \ref{fig:w_actual_table} and \ref{fig:w_table_absolute_difference_actual_vs_expected} shows the objects named ``ship 37'' instead of ``ship 27-46'', ``ship 69'' instead of ``ship 46-91'', and ``ship 92'' instead of ``ship > 91'' from the IAMSAR Manual Volume II \cite{imo2022iamsar}.

Figure \ref{fig:w_table_absolute_difference_actual_vs_expected} shows a table of the absolute difference between \W from the simulation model in Figure@\ref{fig:w_actual_table} and the IAMSAR manual sweep widths for helicopters results. E.g. the absolute difference in \W for detecting a ``ship 92'' at \begin{math}600\end{math} metres altitude greater than \begin{math}37\end{math} kilometres visibility is \begin{math}47.5\end{math} kilometres.

\begin{figure}[h!]
    \centering
    \includegraphics[angle=0, width=\textwidth]{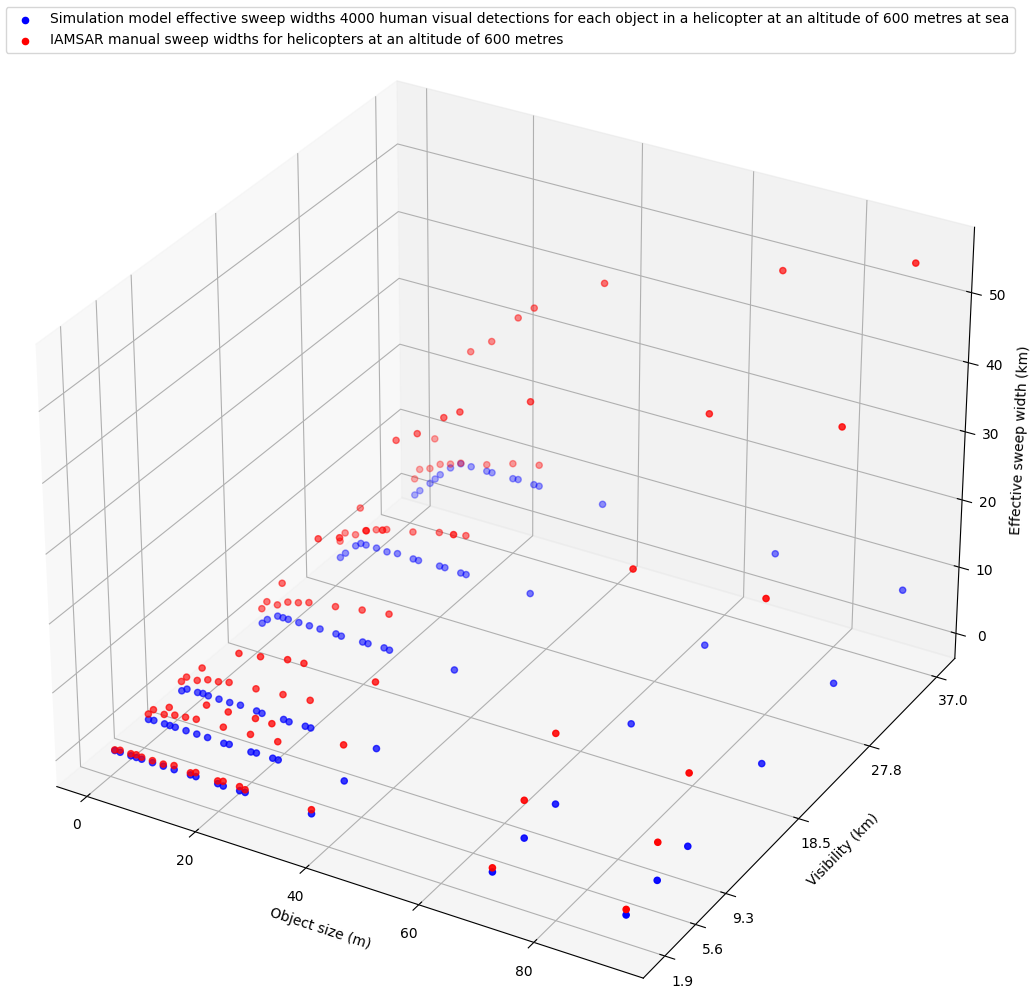}
    \caption{Effective sweep widths scatterplot simulation model 4000 human visual detections for each object in a helicopter at an altitude of 600 metres at sea vs the IAMSAR manual sweep widths for helicopters at an altitude of 600 metres results}
    \label{fig:w_actual_vs_expected_scatterplot}
\end{figure}

Figure \ref{fig:w_actual_vs_expected_scatterplot} shows a three-dimensional scatterplot containing the object size (metres), visibility (kilometres) and \W (kilometres) for the simulation model results vs the expected results. Figure \ref{fig:w_actual_vs_expected_scatterplot} focuses on the altitude of \begin{math}600\end{math} metres case, as other altitudes did not cause a significant difference in their results due to the horizon only being a limiting factor at \begin{math}150\end{math} metres altitude. The figure shows that at the lowest visibility of \begin{math}1.9\end{math} kilometres, the results are somewhat similar, but as the visibility increases, the \W becomes greater. The more significant differences for \W as the visibility increases because the simulation model is limited to \begin{math}4000\end{math} detections. More detections mean more detection opportunities, which will increase the \W.

%% file: sections/conclusions_and_future_work.tex
\label{sec:conc}

In this paper, an initial model of human visual detection in a helicopter at sea using VDM was built to derive a \W table for each object size given the sensor's (human eye) altitude and visibility. The initial model gives an idea of how this can be applied to the camera drone detection case. The model results differ quite from the IAMSAR Manual Volume II \W for helicopters. The reason for such a significant difference is that the model is limited to 4000 detections per object. More detections mean more detection opportunities which will increase \W. The study shows that using a lateral range experiment, you can calculate the detection capability \W for any sensor detecting an object in specific environmental conditions. The lateral range experiment produces a lateral range curve in which the area \W under the curve can be calculated. Future work is to apply the simulation model work done so far to calculate \W in this paper to the camera drone detection case with some improvements. Improvements are split into the sensor (camera drone), object and environmental conditions affecting \W. The sensor should include the camera's angular resolution, human detection performance factors, e.g. fatigue, assuming a human is in the loop, and the sensor moving in real-time. The object should include physical characteristics, e.g. colour. The environmental conditions should include weather, lighting, and obstacles depending on the terrain. E.g. obstacles in the sea would be waves at different heights.

%% file: sections/acknowledgements.tex
This work was supported by the UK Engineering and Physical Sciences Research Council (EPSRC) Doctoral Training Partnership (DTP) with Newcastle University. The authors would like to thank David Perkins and the Centre for Search Research (CfSR), UK registered charity number 1064927.